# Educating Educators to Integrate Inclusive Design Across a 4-Year CS Degree Program

Lara Letaw

Oregon State University, letawl@oregonstate.edu

Rosalinda Garcia

Oregon State University, garciros@oregonstate.edu

Patricia Morreale

Kean University, pmorreal@kean.edu

Gail Verdi

Kean University, gverdi@kean.edu

Heather Garcia

Oregon State University, Heather.Garcia@oregonstate.edu

Geraldine Jimena Noa

Oregon State University, noaguevg@oregonstate.edu

Spencer P. Madsen

Oregon State University, spmadsen21@gmail.com

Maria Jesus Alzugaray-Orellana

Oregon State University, alzugarm@oregonstate.edu

Margaret Burnett

Oregon State University, burnett@oregonstate.edu

How can an entire CS faculty, who together have been teaching the ACM standard CS curricula, shift to teaching elements of inclusive design across a 4-year undergraduate CS program? And will they even want to try? To investigate these questions, we developed an educate-the-educators curriculum to support this shift. The overall goal of the educate-the-educators curriculum was to enable CS faculty to creatively engage with embedding inclusive design into their courses in "minimally invasive" ways. GenderMag, an inclusive design evaluation method, was selected for use. The curriculum targeted the following learning outcomes: to enable CS faculty: (1) to analyze the costs and benefits of integrating inclusive design into their own course(s); (2) to evaluate software using the GenderMag method, and recognize its use to identify meaningful issues in software; (3) to integrate inclusive design into existing course materials with provided resources and collaboration; and (4) to prepare to engage and guide students on learning GenderMag concepts. We conducted a field study over a spring/summer followed by end-of-fall interviews, during which we worked with 18 faculty members to integrate inclusive design into 13 courses. Ten of these faculty then taught 7 of these courses that were on the Fall 2021 schedule, across 16 sections. We present the new educate-the-educators curriculum and report on the faculty's experiences acting upon it over the three-month field study and subsequent interviews. Our results showed that, of the 18 faculty we worked with, 83% chose to modify their courses; by Fall 2021, faculty across all four years of a CS degree program had begun teaching inclusive design concepts. When we followed up with the 10 Fall 2021 faculty, 91% of their reported outcomes indicated that their incorporations of inclusive design concepts into their courses went as well as or better than they had expected.

# 1 INTRODUCTION

Recently, several universities and individual faculty members have begun increasing course coverage of ethical dilemmas that arise in computing professions. Some do so by adding critical thinking to certain CS courses, some by inserting lectures on ethics into their courses, some by adding requirements for more stand-alone ethics courses (e.g., [7,15, 19, 21, 26, 49], and others discussed further in Section 2.2). Despite these efforts, however, researchers have reported that CS students—even those who have become aware of these issues—are not following through to act upon their new awareness (e.g., [20]).

We believe that addressing this problem may require a more comprehensive approach across CS degree programs and that, as part of the approach, students will need hands-on experiences in which they act upon ethical issues they find. Toward this end, we have been working on a new approach with these attributes—in which CS faculty integrate HCI inclusive design methods into multiple courses that gradually build students' skills across all four years in the major, as per Figure 1.

This approach requires buy-in and change by a sizeable number of faculty, coordination across the CS curriculum, and collaboration throughout the faculty to make sure the pieces fit together well. It also requires educating faculty with a variety of specialties (as some faculty in other specialties are not familiar with HCI) to build students' inclusive design skills continually over the four years. These requirements raise numerous challenges. Will faculty want to embark on such a project? Will those without HCI backgrounds be interested and be able to embed inclusive design elements into their "core CS" courses?

In this paper, we present and evaluate an *educate-the-educators curriculum* to support this approach. The curriculum aims to not only teach CS faculty inclusive design concepts, but also teach reasons *why* they might want to teach those concepts to their students and *how* to smoothly embed the concepts into their courses. In support of these goals, the educate-the-educators curriculum aimed to make incorporating and teaching inclusive design low-cost for each faculty member and minimally-invasive into their courses, and to support a pathway through the degree program by which students gradually acquired more and more inclusive design skills, without experiencing repetitive content in multiple courses. We evaluated the curriculum through a three-month field study and subsequent end-of-fall interviews, in which we used the curriculum to teach 18 faculty. Fifteen of the faculty decided to include inclusive design concepts in their courses, which together spanned all four years of an undergraduate CS degree program.

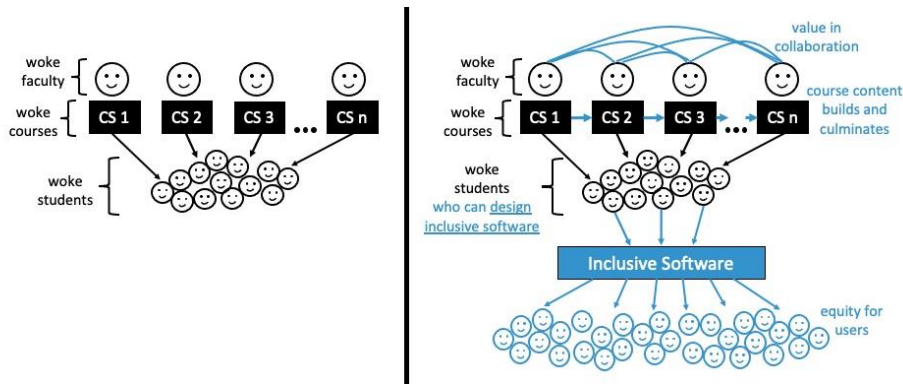

Figure 1: (Left): A pattern common in Responsible CS work. (Right): The approach we were training faculty to carry out across a 4-year undergraduate CS degree program. The differences (**blue**) from many others' approaches to Responsible CS are: (1) faculty collaborate and coordinate on creating (2) gradual, coordinated content across the curriculum to (3) engage students throughout their CS program in an ethical CS practice (creating more inclusive software) that directly impacts their products' users. (Note: "woke" is an informal adjective used in the US to mean "alert to injustice(s) in society".)

Our educate-the-educators curriculum has four elements:

- Curriculum Element #1: *Getting faculty motivated*: Because the success of the project would depend on a sustained and coordinated effort by the faculty, we employed multiple mechanisms to motivate the faculty to engage and stay engaged. At the end of this element, we evaluated Learning Outcome #1: analyzing the cost and benefits of the approach in their own context. We detail these mechanisms and evaluations in Section 4.1.

- Curriculum Element #2: *Teaching faculty inclusive design content*: For faculty to embed bits of inclusive design into core CS courses, they first need to understand inclusive design; in this case, the GenderMag method for inclusive design [8].

We taught faculty how to do GenderMag evaluations using multiple curricular mechanisms and then evaluated Learning Outcome #2: evaluating software using the GenderMag method and recognize its use to identify meaningful issues in software. We detail these mechanisms and evaluations in Section 4.2.

- Curriculum Element #3: *Guiding faculty through embedding inclusive design concepts into their courses*: A critical piece would be whether faculty could and would create integrations of elements of GenderMag into their own courses. We scaffolded their efforts and provided example materials that some faculty at other institutions have used in individual courses. We then evaluated Learning Outcome #3: integrating inclusive design into their existing course materials with provided resources and collaboration. We detail the mechanisms used and evaluation of the learning outcome in Section 4.3.

- Curriculum Element #4: *Developing faculty's pedagogic content knowledge (PCK)*: After faculty had created some of their changes, they began working on how to teach the new content effectively by practicing on each other. We also introduced several PCK findings for how to teach inclusive design effectively. These mechanisms contributed to Learning Outcome #4: preparing to engage and guide students on learning GenderMag concepts. We detail these mechanisms and evaluation in Section 4.4.

The contributions of this paper therefore are:

- An educate-the-educators curriculum to enable faculty to embed inclusive design in a coordinated fashion across a four-year CS program; and

- Results of a three-month field study and end-of-fall interviews investigating faculty's journey starting from early interest in the vision shown in Figure 1 through their fall term in their own classes carrying out that vision.

## 2 BACKGROUND AND RELATED WORK

### 2.1 Inclusive Design with GenderMag

Our educate-the-educators curriculum leverages, as its source of inclusive design content, the GenderMag method's components and foundations. GenderMag [8] is an existing method for avoiding, finding, and fixing inclusivity "bugs" in software. We chose this particular method as it is evidence-based [8] and is used in practice by technologists around the world (e.g., [9, 17, 23, 30, 32, 39, 46]).

At the core of GenderMag are five *facets*, each of which has a range of facet values an individual could have. The five facets are: motivations for using technology; information processing style; computer self-efficacy; learning style[1] (by process or by tinkering); and attitude toward risk. An individual's set of facet values reflects their cognitive styles that contribute to their use of software. GenderMag defines *inclusivity bugs* as failures of a technology product to support the full range of cognitive styles (such as supporting tinkerers but not individuals who prefer understanding a process first before filling in the details). Such barriers are cognitive inclusivity bugs because they disproportionately impact people with particular cognitive styles. They are also gender inclusivity bugs because the facets capture (statistical) gender differences in how people problem-solve [2, 8, 13, 14, 42, 46].

GenderMag uses three personas to bring the facets to life: Abi (Abigail/Abishek), Pat (Patricia/Patrick), and Tim (Timara/Timothy). Abi's and Tim's values for each of these facets lie at opposite ends of the spectrum, and Pat has values within. The Abi persona represents facet values that disproportionately skew towards women, Tim represents facet values that disproportionately skew towards men, and Pat provides a third set of values [8]. The principle behind GenderMag is that technology that *simultaneously* supports all three personas also supports personas with different mixes of Abi's, Tim's, and Pat's facet values. Cognitive styles of the three personas are shown in Figure 2.

The GenderMag method integrates these personas and their facets into a specialized cognitive walkthrough [8, 28]. As with other cognitive walkthroughs [28], a GenderMag walkthrough involves walking through every step of a use-case/scenario and answering questions about each subgoal/action a user "should take" to succeed at the use-case. The GenderMag walkthrough also refers specifically to the persona and facets in each question. The questions are:

- *Before taking any actions*: Will <persona> have this subgoal/take this action? Why/what facets?

---

[1] Here and throughout, "learning style" refers to the GenderMag facet about learning new technologies via process (a top-down style) versus via tinkering (a bottom-up style). This is different from the education community's use of the term "learning styles" referring to learning through different formats (auditory, visual, kinesthetic, etc).

- *After taking the "should take" action*: If <persona> does the right thing, will they know that they did the right thing and are making progress toward their goal? Why/what facets?

Empirical studies have found GenderMag to be effective at identifying inclusivity bugs and at pointing toward effective fixes [8, 9, 17, 23, 32, 39, 46]. However, in the realm of CS education, the only works relating to GenderMag are Oleson et al.'s Action Research investigation into how HCI-oriented faculty teach GenderMag in face-to-face university CS classes [30] and Letaw et al.'s investigation into using portions of GenderMag in two online courses [27]. No prior work has investigated how to enable non-HCI faculty to embed elements of GenderMag into "core CS" across a four-year CS degree program.

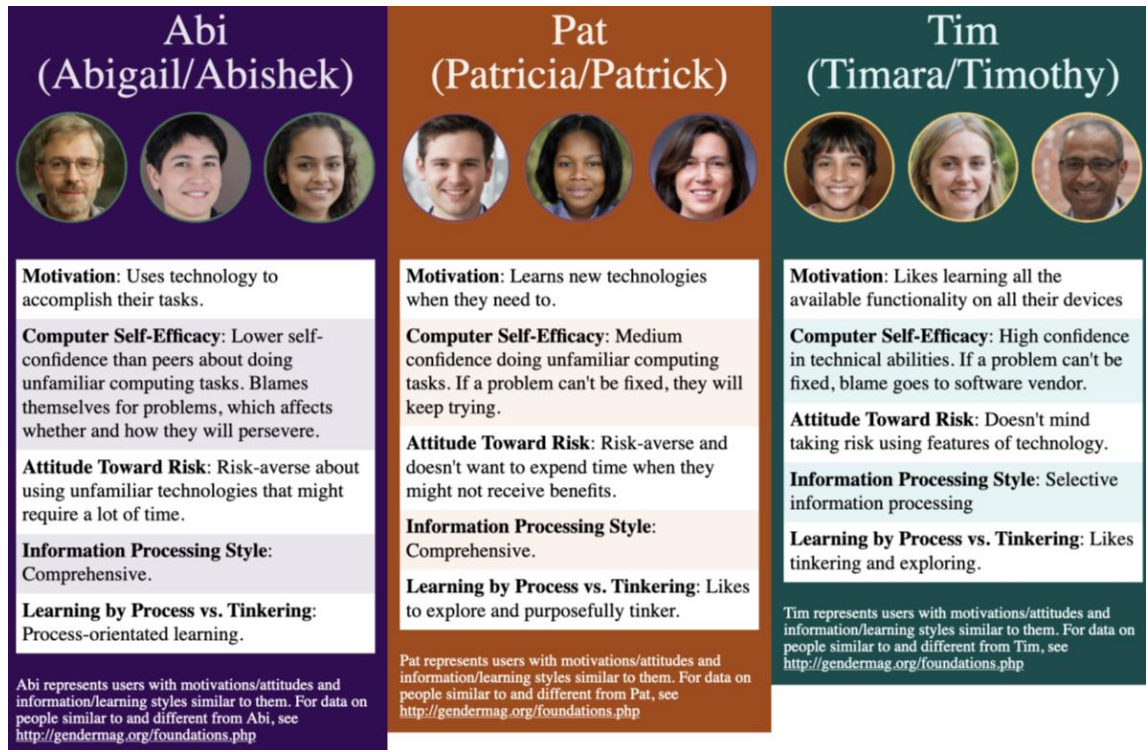

Figure 2: Cognitive styles of the GenderMag personas: Abi, Pat, and Tim

## 2.2 Responsible CS

Our educate-the-educators curriculum aimed to support faculty embarking on an embedded inclusive design approach to CS education, which can be considered a form of Responsible Computer Science. The term "Responsible CS" gained traction with the Mozilla Foundation's Responsible Computer Science Challenge, an effort to teach students mindfulness of ethical problems, respect for stakeholders, and how to make accessible design decisions [15, 29]. "Responsible CS" has similarities to "Critical CS" [26] in that both schools of thought aim to encourage CS students to think critically about ethical behavior and be mindful of the societal implications of their work (e.g., [15, 19, 26, 49]).

Many Responsible CS approaches minimize faculty workload impacts by not directly involving CS faculty. For example, one project at Harvard successfully integrated ethical reasoning throughout mainstream CS courses by bringing philosophy graduate students and postdocs into the CS courses [21]. Another project at Brown infused social justice and responsible computing strategies throughout their CS curriculum through the work of undergraduate teaching assistants [15]. A third project created a repository of ethics modules that faculty could choose from but weren't involved in creating [7]. Similar examples can be found in [3, 11, 16, 36, 37, 38] as well as examples of approaches with separate ethics courses (e.g., [10, 18]). Because approaches like these do not directly involve CS faculty, they may not need an educate-the-educators curriculum.

Another common thread among many Responsible CS and Critical CS approaches is an emphasis on building students' skills to reason about ethical *problems* arising in CS, but not explicitly skill-building their abilities to solve such problems. For example, at one university, classes discuss targeted advertising, bias, disinformation, and other ethics-related topics with the aim of learning and reflecting upon these issues [15]. Similarly, another university's Responsible CS initiative goal was to teach students to be conscious of ethical problems while reasoning and communicating about them [21]. However, one recent study suggests that approaches like these may not be enough to make a difference in CS students' ethical *behaviors* [20]. This study showed that even when students were aware of ethical issues, they often did not act upon them ethically. This result suggests that CS faculty may need to teach students new skills for *solving* ethical problems.

The approach we present in this paper differs from both of the above approaches. First, it emphasizes each CS faculty member having ownership of and responsibility for their own class's Responsible CS elements. Second, it does so by educating the faculty on how to teach skills for *solving* a class of ethical problems.

## 2.3 Inclusive Design Education

Our educate-the-educators curriculum aims to provide CS faculty with the background and skills to embed inclusive design education into their own "mainstream" CS classes. A number of studies suggest that enabling CS faculty to do so can improve inclusivity in CS classes. For example, the DO-IT project created a web development course that integrates accessibility and universal design [1]. To increase feelings of inclusion for both women students and students with disabilities, Blaser et al. proposed including universal design principles in engineering courses to prepare future engineers better and improve representations of disabled users and engineers [4]. Blaser et al.'s research rests on prior investigations of what diverse students value in their courses and jobs, reporting that women in engineering often value contribution to society more than men do, which suggests that women may be drawn to inclusive and universal design (e.g., [22]). Similarly, Izzo et al. have found that teaching universal design in college courses in order to include people with disabilities helps both students and instructors to improve accessibility, awareness, and instructional flexibility [24]. Oleson et al.'s recent study shows the need for such integrations [31]. In this study, HCI students themselves reported lacking ability to design for diverse populations or to guard against biasing their designs, and 54% and 49%, respectively, of HCI faculty reported occurrences of these inabilities among HCI students.

Closest to our approach to embedding inclusive design into mainstream CS classes is work from Waller et al. and Putnam et al. Waller et al. experimented with integrating accessibility across a curriculum [47], although their approach is less minimally invasive than needed for our goals. Similarly, Putnam et al. suggest guidelines for achieving this goal in the accessibility domain [34], and we apply their recommendations to our strategy. For example, to enable diverse faculty to include accessibility-related concepts in their courses, Putnam recommends: (1) a combination of multiple unique learning experiences (and pointers to key open problems); (2) resources to enable a broad set of CS faculty who are not inclusion specialists to integrate teaching inclusion (for Putnam, accessibility); and (3) helping CS faculty know how to evaluate the effectiveness of curriculum that include these aspects of diversity [34]. Our educate-the-educators curriculum is built on all three of these recommendations.

In prior years, individual faculty at various institutions have also included elements of GenderMag (e.g., the GenderMag personas or the method itself) in their courses, especially HCI and Software Engineering courses (e.g., [27, 30]). However, prior work has not investigated how to *enable a broad spectrum* of faculty, including those not trained in HCI, to continually add to students' ability to build inclusive software *across a coordinated, 4-year curriculum*.

## 2.4 Educating Educators to Teach Inclusive Design Content

Our educate-the-educators curriculum also draws upon three foundations into how to support faculty's teaching: Communities of Practice, the Training of Trainers model, and Pedagogic Content Knowledge (PCK).

A Community of Practice is a learning community that often engages in informal learning and professional networking [48]. Scholars have applied the concept to different types of learning communities such as professional learning communities [25], professional learning networks [44], and faculty inquiry groups [5]. Community of Practice approaches recognize that professional development among faculty is a social activity and that communication among participants is key. A Community of Practice has three main components: a shared area of interest (the domain), a group of people who engage and share knowledge (the community), and a shared collection of resources (the practices) [48]. In our setting, the community's shared interest (domain) was embedding inclusive design across their CS/IT curriculum, the people (community) were the faculty who were working toward this shared goal, and the shared collection of resources was a wiki of teaching resources and curricular materials which we seeded and the faculty eventually contributed to. Our educate-the-educators curriculum used active learning exercises and small-group hands-on work to draw the faculty into a Community of Practice.

The active, hands-on aspects of Communities of Practice also helped to align our educate-the-educators curriculum with the Training of Trainers (ToT) model [6, 12, 33]. ToT programs are widely used in medical and health settings and have also been used in other educational settings [6]. For example, in a higher education context, Peterson used a ToT approach to teach students about civic engagement, so that they may take leadership roles moderating or facilitating public deliberation [33].

Our educate-the-educators curriculum draws upon multiple ToT properties. Two that are essential in the ToT model are that it emphasizes (1) teaching participants, the relevant *content* (2) using *evidence-based* approaches for teaching the content to others. These properties of ToT informed our curriculum: we included teaching educators GenderMag content and how to apply GenderMag, which is itself an evidence-based approach [8]; and we followed evidence-based approaches to *teaching* GenderMag (e.g., [30]). Other components of the ToT method that we integrated into our educate-the-educators curriculum included modeling and skill practice, active learning activities, opportunities for feedback, and follow-up support. For example, we walked the faculty through applying GenderMag in one of our sessions and then the faculty applied the method themselves, working in small groups. This skill practice and active learning with peers then helped to scaffold their lesson planning. Action planning is another important property of the ToT model [6] that we integrated into our educate-the-educators curriculum. In our case, participants created or revised their own course lesson plans to integrate applicable portions of GenderMag content, receiving feedback from the trainers and their peers along the way.

To enable the faculty to embed inclusive design curricular changes in their course, we also drew upon research into applicable Pedagogic Content Knowledge (PCK) [40]. PCK is the intersection of pedagogical knowledge (background in effective teaching techniques and practices) and content knowledge (background in the subject being taught) that enables faculty to teach *particular* content. PCK is specific to the topic at hand and to the audience [45], so we supported the faculty's curricular changes by building upon Oleson et al.'s investigation into PCK enabling faculty to teach inclusive design skills using GenderMag [30], a point we will expand upon in Section 4.4.

## 3 ENABLING FACULTY TO EMBED INCLUSIVE DESIGN, AND EVALUATION METHODOLOGY

We built upon the above foundations to implement the four curriculum elements we briefly described in the Introduction—(1) a curriculum element to motivate the faculty; (2) a curriculum element to teach faculty concretely how to do inclusive design with GenderMag; (3) a curriculum element to help faculty create suitable integrations for their own courses; and (4) a curriculum element to help faculty develop their pedagogic content knowledge. All of the materials we used to implement these curriculum elements are included in the Supplemental Documents accompanying this paper. In the next subsections, we explain our field study methodology, and in the next section we present the results of the field study in the context of detailing each of the curriculum elements being evaluated.

### 3.1 Field Study Span and Context

To investigate the efficacy of our educate-the-educators curriculum, we conducted a field study over a three-month period from early May through late July. We then followed up with the faculty at the end of the subsequent fall term (November and December).

The field study investigated the on-the-job endeavors of faculty to adjust their own courses to integrate inclusive design concepts in time for fall term. They had full ownership of how they adjusted their own courses. In addition, the CS faculty participants' courses together had to fulfill the department's four-year, whole-curriculum goal of embedding inclusive design. Thus, the courses each year needed to build upon the material taught in previous courses, without counterproductive overlaps or gaps in needed inclusive design prerequisites.

The faculty's on-the-job environment during the three-month field study was virtual due to COVID lockdowns. Thus, interactions consisted of shared/exchanged documents, emails, online resources, document exchanges, and a Zoom virtual meeting room where participants and facilitators could interact and show what was happening on their screens. The educate-the-educators curriculum was offered to the faculty via: (1) a 12-hour interactive workshop series split over multiple days and offered twice—one on two consecutive Saturdays in May for 6 hours per day, and the other on three consecutive weekdays for 4 hours per day (roughly same number of participants in each); (2) a set of resources including an online community of shared examples and materials that other faculty teaching GenderMag concepts had used; (3) feedback to faculty on their work over the ensuing weeks; and (4) occasional emails answering individual questions and providing updates to the online community's shared materials. In contrast to the online environment during the field study, all faculty's fall classes were entirely in-person.

## 3.2 Participating Faculty

Of the 18 participants, 16 were CS faculty at a single university engaged in the across-the-degree-program effort described earlier. These 16 taught CS and IT courses at a public regional Hispanic Serving Institution. Class size in the CS/IT department is relatively small (20-25 students) and faculty are rewarded for emphasizing quality teaching. CS faculty cover a BS in Computer Science and a BS in Information Technology program. The remaining two participants were a non-CS (education) faculty member at the same university, and a non-CS (electrical and computer engineering) faculty member at a different public university. These two non-CS participants were engaged in changing only one course each without a need to coordinate with other courses.

We recruited the faculty through the CS department chair. The chair first canvassed the department faculty and they expressed interest in a coordinated adjustment to the four-year CS degree. When we then offered the opportunity for individual faculty to participate with their own courses for a modest $500 summer stipend, 15 opted in. Three others eventually joined who had heard about it through word-of-mouth (without the stipend). Participant demographics and course coverage are shown in Table 1.

Table 1. (Left): Faculty participants. (Right): Courses of the 18 faculty participants that they identified for potential inclusion of GenderMag concepts during academic year 2021-22. (Column does not add up to 18 faculty because some faculty were involved in more than 1 course.) Mobile Apps was not part of our original plan, but the faculty member took the initiative to add GenderMag to that course.

| Faculty teaching load | Men | Women |
|---|---|---|
| CS Full time (3-5 sections/semester) | 5 | 5 |
| CS Part-time (Adjunct, 1-2 sec./sem.) | 4 | 2 |
| Total CS faculty | 9 | 7 |
| ECE faculty (different university) | 0 | 1 |
| Education faculty (different dept.) | 0 | 1 |
| Grand Total | 9 | 9 |

| Undergraduate CS/IT courses covered | # of faculty |
|---|---|
| CS0 (Intro to Programming) | 3 |
| CS1 (OOP) | 3 |
| CS2 (Data Structures) | 1 |
| Object Oriented Design | 2 |
| Web Design | 2 |
| Mobile Apps | 1 |
| Human Computer Interaction | 2 |
| Software Engineering | 1 |
| Project Management | 1 |
| Databases | 2 |
| Senior Capstone | 2 |
| | |
| Intro to Engr. (different university) | 1 |
| Education course (different dept.) | 1 |
| Total: 13 courses | 18 faculty |

## 3.3 Activities

Over the three months in the field study, the faculty participated in the 21 activities in Table 2. Before beginning the activities in the table, we reviewed the IRB-approved informed consent document with the participants to gather their consent for our data collection. At the end of the three-month period, faculty whose courses were on the fall term schedule took their curricular changes into the classroom; we then interviewed those faculty members at the end of fall term.

## 3.4 Data

We collected faculty data multiple times throughout the field study (Table 2), via questionnaire responses, workshop recordings, observation notes, faculty-created artifacts, and interviews. In the questionnaires, faculty reflected on their learning, interactions with peers, and their GenderMag integration efforts. The questionnaire data were qualitative with text-entry, Likert scale, and multiple-choice questions. The faculty-created artifacts came from activities faculty performed during the workshop, emails with questions, and course materials the faculty developed over the summer. Faculty iteratively improved these materials over the field study period, receiving two rounds of feedback from facilitators (detailed in Section 4). We collected their versions from the workshop's initial brainstorming, their first submitted versions after the workshop, and then after revisions of their second submission. Finally, at the end of fall term, we conducted and recorded a 30-minute semi-structured interview with each participant who taught one of the updated classes during fall term. These interviews allowed us to compare faculty's pre-classroom expectations with their actual classroom experiences. Interview questions are given in the Supplemental Documents.

We qualitatively coded the faculty interview data. We derived the codebook from the results of the first three months, which will be shown in Tables 3, 4, 5, and 7. Of the fifteen total rows, we used the nine that were relevant to the interview content. To code the data, we first segmented each interview by question. Then, two researchers independently coded 21% of the data, with 80% agreement (Jaccard method) [41]. Given this level of agreement, the same researchers divided up coding the remaining data. The full codebook can be found in the Supplemental Documents.

We used these data to evaluate the educate-the-educators curriculum learning outcomes as follows. We evaluated each curriculum element as a whole using the percentage achieved *collectively* by the faculty. In addition, we evaluated how many *individual* faculty members achieved the learning outcome using 60% averages as our threshold value (because 60% is the passing grade threshold in 90/80/70/60 grading). In practice, this became a 100% threshold for learning outcomes with two measures, and a 67% threshold for learning outcomes with three measures.

Table 2: Faculty engaged in the 21 activities shown, producing the data shown at the end of each group. (Time lengths approximate.)

| When? | Presentation/activity and data | Why included |
|---|---|---|
| Workshop Day 1: The 'content' part of PCK | Intro to inclusive design and our objectives (1hr) | Initial context |
| | What is GenderMag, how does it work, and who else uses it? (15m) | Brief introduction to GenderMag |
| | Cognitive styles activity (25m) | Icebreaker / core GenderMag concept / broadly-applicable activity |
| | GenderMag method lecture (30m) | Introduce inclusive design and the GenderMag method |
| | GenderMag active learning (2hr) | Faculty learn to do GenderMag method |
| | Debrief + feedback questionnaire | Collect response/improvement data |
| Workshop Day 2: The curriculum part | Intro to Starter Packs (5m) | Faculty get content ideas for materials integration |
| | Experiences teaching GenderMag (20m) | Faculty get a sense of what to expect their first time integrating / teaching GenderMag |
| | Backward Design Template (5m) | Faculty get process ideas for materials integration |
| | Hands-on: Integrate GenderMag into your course + practice teaching (2hr) | Faculty get time to develop materials, work collaboratively, practice teaching, and get feedback from experts on their curriculum |
| | Debrief + feedback questionnaire | Collect response/improvement data |
| Workshop Day 3: Integrating PCK | Pedagogical Content Knowledge (PCK) Intro (40m) | Faculty learn effective ways of teaching inclusive design content |
| | Hands-on: Modifying materials + re-teaching (2hr) | Faculty practice teaching the content and get a sense of what students might experience |
| | Discussion (1hr) | Wrap up |
| | Debrief + feedback questionnaire | Collect response/improvement data |
| Summer: Material Submissions & Revisions | Pre-materials submission (Deadline approximately 30 days after Workshop | Collect baseline course materials (pre-GenderMag) *Faculty may not have known which materials would be adapted prior to the workshop, thus these were requested post-workshop |
| | Material submission #1 (Pre-revisions) (Deadline approximately 30 days after Workshop) | Collect first draft GenderMag-integrated course materials *Materials were sent back with feedback 12 days after they were turned in |
| | Material submission #2 (Post revision) (Deadline 22 days after submission #1) | Collect GenderMag-integrated course materials after revisions *Some materials were sent back with additional feedback 6 days after they were turned in |
| Summer: Surveys | Follow-up questionnaire #1 (Approximately 3-4 weeks after the workshop) | Collect response/improvement data |
| | Follow-up questionnaire #2 (Approximately 30 days after follow-up #1) | Collect response/improvement data |
| End of Fall Term | Individual interviews (30m) | Collect teaching experiences from fall faculty |

## 4 EDUCATE-THE-EDUCATORS CURRICULUM AND FIELD STUDY RESULTS

This section details the elements of our educate-the-educators curriculum, how we conveyed the curriculum during the workshop and post-workshop follow-ups, how faculty responded to the curriculum, and what happened when faculty applied what they learned by teaching GenderMag concepts during Fall 2021.

## 4.1 Curriculum Element #1: Getting Faculty Motivated

Motivating the faculty was critical to the project's success because, if the faculty were not motivated, they would be unlikely to be effective in contributing to the coordinated four-year effort. Thus, our objective with this element was Learning Outcome #1: enabling faculty to analyze the costs and benefits of integrating inclusive design into their own course(s). We implemented Curriculum Element #1 through the following five mechanisms:

---

**Curriculum Element #1: Motivating the faculty**

- (1) Appeal to costs/benefits/rewards as per the faculty reward system
- (2) Explain the coordinated "big picture"
- (3) Emphasize each faculty member's control over their own courses
- (4) Explain the equity and inclusion benefits
- (5) Provide data on prior student outcomes and experiences

---

Regarding mechanism (1), this university's reward criteria were quite conducive to the costs/benefits/reward aspect of this curriculum element. This university's reward system emphasizes effective teaching as its top criterion for faculty retention and encourages new teaching techniques. Participating in this program was a way to receive credit for doing so:

(University's faculty retention criteria): "List any new teaching materials, teaching techniques, etc., …"

(CS Dept Chair, interview): "Professional development is encouraged and must be documented."

For mechanisms (2) and (3), we presented the "big picture" to show not only the coordinated whole but also to appeal to some faculty's enjoyment of a collaborative activity (Figure 3).

P12 (Day 2 Feedback): "<Excited about> seeing the connections between the classes."

Figure 3: The "big picture" showed to faculty to guide their efforts and emphasize the need for coordination and lack of overlap (Left): Excerpt from course list for the first two years, showing how the pieces fit together and the minimal classtime expected. For example, for the Intro CS course, we suggested adding only 1/2 new lecture, plus updating an exercise and assignments. (Right): The second two years. HCI1 and SE1 were the only courses for which we recommended significant lecture/classtime additions.

We drew upon a Community of Practice approach (as explained in Section 2.4) by emphasizing that each faculty member has an important role to play in the delivery of the curriculum to students and that effective collaboration and coordination would bring about better outcomes. At the same time, we also emphasized to faculty that they were not being asked to hand over control of their course content. It was important that they see the value in the community efforts, yet not feel constrained by them. To support this, the presentation offered *suggestions* on how to fulfill each learning outcome, not *requirements* (mechanism 3). In fact, faculty would need to be creative to make their integration authentic.

As Figure 3 shows, the presentation also provided some cost/benefit information (mechanism 1), showing the minimal classtime the new material would occupy and the resources faculty could use/customize if desired (e.g., sample lectures, sample active learning exercises, etc.). To further emphasize low time cost and community support, we connected them to a Slack channel for sharing emerging materials with each other.

For mechanisms (4) and (5), we turned to data from other CS education projects. For example, we presented data on students' responses, success rates, and diversity/equity/inclusion results from other studies on faculty who had taught with GenderMag [27, 30]. These data were compelling to some faculty:

P13 (Day 2 Feedback): "Excited to see how it influences the <students'> project teams (and how they work together)"

P17 (Day 2 Feedback): "Great to hear that students felt more inclusive and learned about their own processing style."

At first, some participants anticipated needing to spend significant time on the integration efforts:

P15 (Day 3 Feedback): "I think I need to update a lot of assignments…"

However, by Day 3 all reporting participants had converged to anticipating a light or medium burden and some participants commented that the work was important:

P06 (Day 3 Feedback): "…it is not a burden to include <GenderMag in courses> and learn about how to be more inclusive."

By the end of the three-month period, 14/17 of reporting participants reached the learning outcome threshold of 60% (practical threshold: 67% since there were three measures), as shown in Table 3.

Table 3: Faculty responses relating to Curriculum Element #1, as of follow-up questionnaire #2. Note that after day 1, P08 withdrew due to illness. Overall, the majority of participants had noted a benefit of integrating portions of GenderMag into their course by the time we collected follow-up questionnaire #2.

✓: yes/agree, ✗: no/disagree , -: neither agree nor disagree , n/a: could not respond (e.g., did not do the questionnaire)

| | P01 | P02 | P03 | P04 | P05 | P06 | P07 | P08 | P09 | P10 | P11 | P12 | P13 | P14 | P15 | P16 | P17 | P18 | Total |
|---|---|---|---|---|---|---|---|---|---|---|---|---|---|---|---|---|---|---|---|
| Burden to integrate light? (Day 3) | ✓ | - | - | ✓ | - | ✓ | ✓ | n/a | - | - | - | - | - | - | - | ✓ | ✓ | n/a | 6/16 |
| Material relevant to your students? (Follow-up #2) | ✓ | ✓ | - | - | ✓ | ✓ | ✓ | n/a | ✓ | ✓ | ✓ | ✓ | ✓ | ✓ | ✓ | ✓ | ✓ | n/a | 14/16 |
| Benefit of incorporating into course? (By end) | ✓ | ✓ | ✓ | ✓ | ✓ | ✓ | ✓ | ✗ | ✓ | ✓ | ✗ | ✓ | ✓ | ✓ | ✓ | ✓ | ✓ | n/a | 15/17 |
| | | | | | | | | | | | | | | | | Overall success rate: 35 ✓s out of 49 possible (71%) | | | |

## 4.2 Curriculum Element #2: Teaching Faculty GenderMag Content

For faculty to teach content on inclusive design, they would need to gain expertise in the content area, so Curriculum Element #2 taught inclusive design concretely in the form of the GenderMag method (Section 2.1). The associated learning outcome, Learning Outcome #2, was enabling faculty to evaluate software using the GenderMag method and recognize its use to identify meaningful issues in software. We aligned this element with several steps from the Training of Trainers research (as described in Section 2.4), particularly providing information on how people learn through the GenderMag cognitive styles and offering opportunities for practicing new skills through active learning [12]. Our mechanisms for Curriculum Element #2 were as follows:

> **Curriculum Element #2: GenderMag content**
> - (1) Cognitive styles sharing activity
> - (2) GenderMag lecture
> - (3) GenderMag active learning activity (learning by doing)

Even though many faculty would be teaching only a portion of the GenderMag method, the curriculum element included the full GenderMag method. One reason for including the full GenderMag method was to illustrate to faculty the proficiency we wanted students to acquire by the end of their degree; another was to model to faculty how we teach the components of the method (part of Curriculum Element #4); and a third reason was that teaching a specific method provided a concrete, hands-on way to introduce faculty to inclusive design.

The first mechanism by which we taught this content was a cognitive styles sharing activity. Faculty were asked to share their facet values by placing a marker with their name on the scale as shown in Figure 4. These facet values were defined earlier in Figure 2, which we also presented to the faculty.

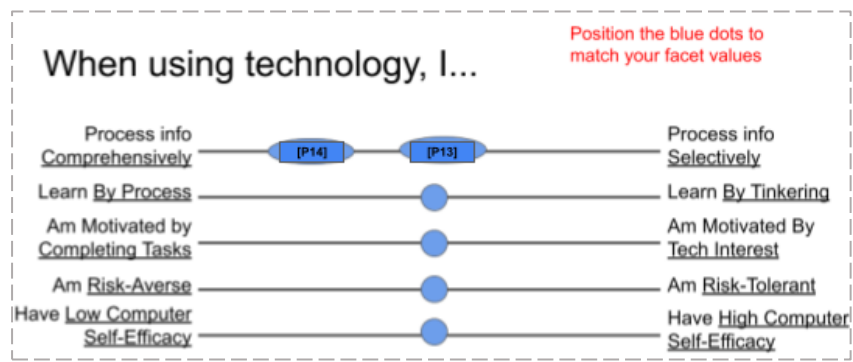

Figure 4: An example of part of the cognitive style sharing activity from two of the faculty participants who learned they process information differently. This was a PowerPoint-based version of the activity from the June workshop (which worked better than a Jamboard-based version in the May workshop).

The activity fulfilled several functions. It served as a community builder as faculty came to know how their peers tend to problem-solve. At the same time, it offered faculty insight into how users learn to use technology in different ways. It also introduced central concepts of GenderMag and was an activity suitable for any CS/IT course requiring groupwork by the students. The activity resulted in a rich whole-group discussion, with faculty engagement becoming more personal. For example, faculty were able to connect the activity's implications to their own teaching styles and to their students.

> P07 (Day 1 transcript): "…I think I am more like Tim…I need to feel that I am free to make errors. This is also a part of my teaching style."
>
> P14 (Day 1 transcript): "…in the context of teaching, students would be all over the spectrum…"
>
> P13 (Day 1 transcript): "…it would be interesting to see…<how students form groups> in terms of…problem solving styles and so forth."
>
> P12 (Day 1 transcript): "It also helps you spot the conflicts in the groups…the ones who want to complete tasks versus the ones who are tech interested…you get a boiling point at some point time."

Building upon the faculty's understanding of cognitive styles, mechanism (2) presented the full GenderMag method as a 30-minute lecture. The lecture introduced inclusive design through the GenderMag method—in ways that did not require an HCI background, so it would serve as appropriate introductory material for both faculty and students. In this introductory segment, the facilitators modeled the application of the GenderMag method to provide a concrete example faculty could later use to engage their students. The slide deck, derived from the GenderMag repository[2], is provided in the Supplemental Documents. The lecture was followed by a two-hour active learning component, in which faculty split into small groups to use GenderMag to evaluate and improve a software product's inclusivity.

The activity involved each small group "walking" through a use-case of a software design we had provided, at each step evaluating the usability experience a GenderMag persona might have. Faculty considered a use case for Canvas, a popular education software platform, from the perspective of the Abi persona. With this perspective, they answered questions about Abi's use of the software as described in Section 2.1. The activity materials are given in the Supplemental Documents. We coached faculty on this activity and regularly brought them back together for sharing their results and large-group discussions. However, most of the time they were learning without us, mirroring an active-learning in-class activity they might use in their own classes. By the end of the activity, they had gained skills in locating "inclusivity bugs" in software and were suggesting the fixes.

> P05+P06+P08 (written portion of GenderMag evaluation, Day 1): "There's no indication of progress/process."
> (inclusivity bug for Abi, relating to Computer Self-Efficacy and Learning Style)
>
> P01+P02 (written portion of GenderMag evaluation, Day 1): "…the association between the account and the actual video is not clear and <Abi is not> a risk taker…Maybe show the video list. Then show the account…" (bias against Abi found for Attitude Towards Risk and Learning Style facets)

By the end of Day 1, 14/17 (82%) of reporting faculty members reached the learning outcome threshold of 60% (practical threshold, given two measures: 100%) with 14/17 who reported being able to do a GenderMag evaluation and 16/17 (94%) who said that using GenderMag had revealed meaningful issues in the software they had evaluated (Table 4).

Table 4: Curriculum Element #2 results, from faculty responses to Day 1 feedback questionnaire. 16/17 of the faculty met the learning outcome threshold. ✓,✗,-,n/a: same as Table 3.

| | P01 | P02 | P03 | P04 | P05 | P06 | P07 | P08 | P09 | P10 | P11 | P12 | P13 | P14 | P15 | P16 | P17 | P18 | Total |
|---|---|---|---|---|---|---|---|---|---|---|---|---|---|---|---|---|---|---|---|
| Can do GenderMag eval? | ✓ | ✓ | ✓ | ✓ | ✓ | ✓ | ✓ | ✗ | ✓ | - | ✓ | ✓ | ✓ | ✓ | ✓ | - | ✓ | n/a | 14/17 |
| GenderMag found meaningful issues? | ✓ | ✓ | ✓ | ✓ | ✓ | ✓ | ✓ | ✗ | ✓ | ✓ | ✓ | ✓ | ✓ | ✓ | ✓ | ✓ | ✓ | n/a | 16/17 |
| | | | | | | | | | | | | | | | | | | Overall success rate: | 30/34 (88%) |

## 4.3 Curriculum Element #3: Guiding Faculty Through Embedding GenderMag Concepts into Courses

In Curriculum Element #3, faculty needed to act upon what they had learned—i.e., decide what portions of inclusive design content they wanted to use in their own courses and how. To support their efforts, this curriculum element used four mechanisms which leveraged a Training of Trainers approach through an emphasis on practice and feedback, action planning with backward design, and multiple support opportunities for faculty [6, 12, 33]. We also continued to foster a Community of Practice through an ongoing emphasis on peer collaboration and support. Thus, the desired Learning Outcome #3 was for faculty to be able to integrate inclusive design into existing course materials with provided resources and collaboration.

---
**Curriculum Element #3**
- (1) Process ideas (backward design, starter packs, and example integration)
- (2) Content ideas (online community)
- (3) Material creation with coaching and collaboration
- (4) Material submissions and feedback

---

Mechanism (1) scaffolded the faculty's creative efforts with materials defining a *process* they could follow, including examples of following it. The process, backward design [35], is a well-known process for curriculum creation. It starts by considering how to assess whether students have achieved the desired learning outcome, then considers assignments that could be assessed in that way, and finally designs portions of the course that would enable students to succeed at such an assignment. For example, Figure 5 shows P06's use of this process.

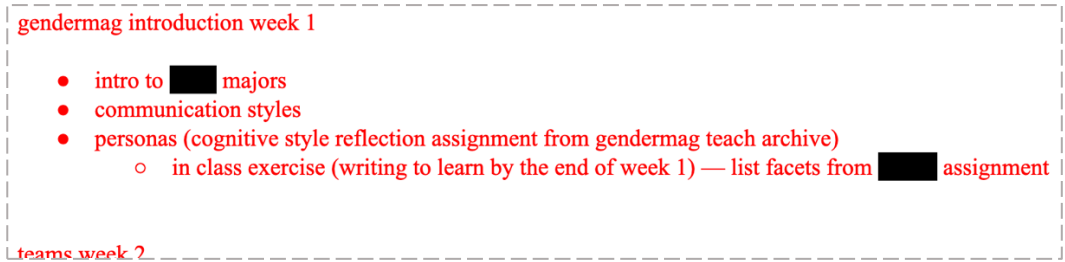

Figure 5: Excerpt from P06's three-week curriculum plan created using the backward design template.

To further support faculty's action planning, in mechanism (2) we provided "starter packs": templates for each year of a 4-year CS degree, with fill-in stages for backward design and suggested materials housed in an online community for faculty who teach with GenderMag [43]. The online community content includes lecture slides, homeworks, readings, in-class activities, and exam questions that participants could reuse or adapt. As one example, nine faculty incorporated a GenderMag personas graphic (shown previously in Figure 2) into their course materials. In the Day 2 feedback, 14/16 (88%) of reporting faculty responded that the starter packs were useful and 8/16 (50%) responded, when we followed up after the workshop, that they used the online community frequently.

Mechanism (3) encouraged collaboration as the faculty began their course changes. Faculty joined Zoom breakout rooms of two to four faculty teaching similar courses. Facilitators also visited each room from time to time to offer additional coaching if needed. Most participants' responses to these mechanisms were very positive.

P06 (Day 2 feedback): "<Participant's> revision/additions to my course outline were amazing."

P16 (Day 3 feedback): "The breakout rooms were very collaborative and lots of interesting ideas."

P15 (Day 2 feedback): "…break out group <made me feel excited about integrating GenderMag>"

P02 (Follow-Up #2): "<P01> spoke of the Facets details …helped me to see the whole picture of this research project and helped me to navigate through the course material."

P07 (Day 2 Feedback): "Many of the recommendations <facilitator> made for <participant> were helpful for me to think through how I can utilize this to help my students understand how cognitive styles impact the way students learn."

That said, some of their collaborations had rocky starts. After the first day of the workshop, two participants reported a "somewhat poor" experience working with peers and three reported it went only "somewhat well". Reasons given were interpersonal, technological, and task-related.

P06 (Day 1 feedback; peer work went somewhat poorly): "…I ended up trying to get folks involved and then stepped back because I do not like being in that role continuously."

P05 (Day 1 feedback; peer work went somewhat well): "Lots of confusion due to screens not being shared for long…"

P07 (Day 1 feedback; peer work went somewhat well): "My peers were great, but I think we are all figuring out how to complete the tasks."

However, by the end of the workshop, all reports were that peer work went well, as Table 5 will show at the end of this section. Mechanism (4) added a feedback loop over the next few weeks in which participants turned in their course materials and received feedback. Of the 18 participants, 16 turned in materials for feedback (e.g., Figure 6). Two participants' materials were approved without further changes. However, the remaining materials (14 participants) required some iteration. For example, some participants added too much GenderMag material too early in the four-year sequence, and others too little GenderMag where it was needed in the four-year sequence. Facilitators provided feedback for these 14.

After getting feedback, most participants continued but two withdrew. These two participants had been collaborating on materials for backend-focused courses. But in our feedback, we had asked them to apply inclusive design content to a different part of an assignment: a GUI implementation activity instead of a backend technical learning activity. In retrospect, we believe our feedback may have undercut our own goals because the faculty responded:

P04 (Email after first iteration of feedback): "We think…good candidate courses for GenderMag should be GUI-related…"

Our feedback might have been more successful had it supported integration into the backend content, which would have been both feasible and appropriate for the course. The participants' decision to withdraw also reinforced Curriculum Element #1's third mechanism: Emphasize each faculty member's control over their own courses. The two participants might have felt we were reducing their control over their own courses.

The remaining faculty who needed revisions submitted a second round of materials. The facilitators fully approved 9/14 (64%) without further changes; the rest received suggestions for minor changes.

From the original group of 18 faculty, 15 decided to teach their courses with GenderMag content embedded.

Three examples of their GenderMag-integrated course materials are included here. First, Figure 6 shows P10's and P11's development of an assignment for an introductory programming course, which incorporates an informal evaluation of software using the GenderMag personas. Second, Figure 7 shows an excerpt from activity created by P06 and P07 which introduces the GenderMag facets for an introductory computing course and for P07's education course. Finally, Figure 8 shows a project created by P18 which includes the use of the full GenderMag method.

By the end of this curriculum element, 15/18 (83%) of faculty reached the learning outcome threshold of 60% of the measures in Table 5.

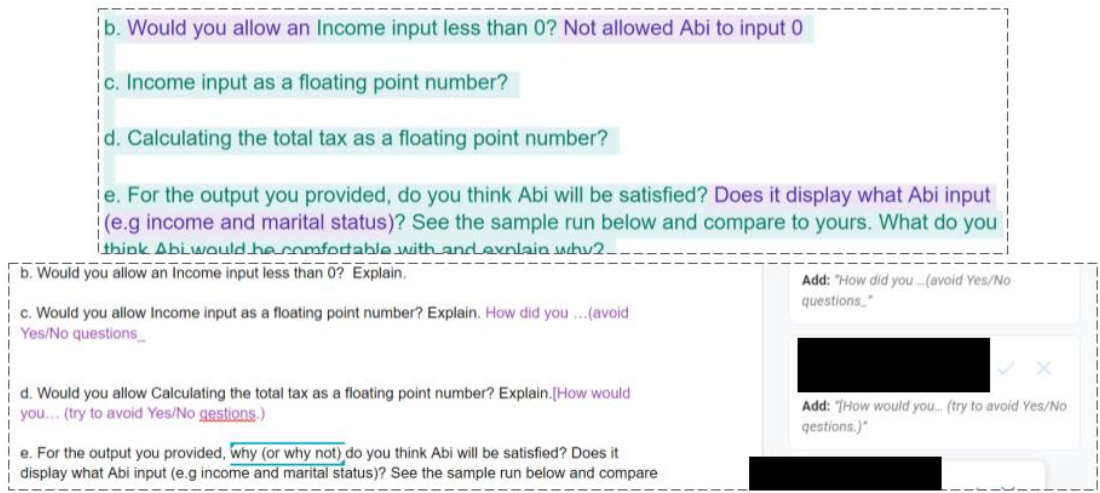

Figure 6: (Top): Changes made by P10 (purple) and P11 (green) as they collaborated on curricular materials during the workshop. (Bottom): After submitting a more polished version of the materials, facilitators responded with specific suggestions for improvements.

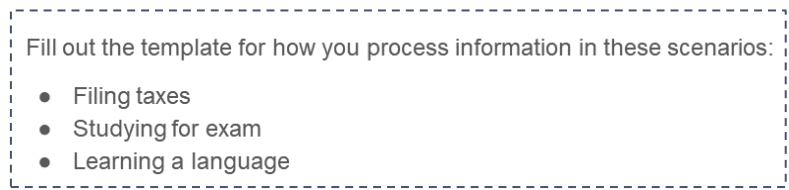

Figure 7: In-class activity by P06 and P07 for students to identify their own facet values. The template referenced is similar to that used in the workshop, shown earlier in Figure 4.

Each individual team member should submit this homework. Include 3 items in one file:
1. Answer all the questions on this homework.
2. The persona that you created (not your team!)
3. The cognitive walkthrough that your team developed. Be sure to identify your contribution to the cognitive walkthrough.

Figure 8: (Top): An excerpt from the final project given by P18 in which students were asked to complete a cognitive walkthrough on their team's software. (Bottom): Part of the customizable persona provided for students to use in their cognitive walkthrough.

Table 5: Curriculum Element #3 results, from faculty questionnaire responses and the curriculum materials they created. Overall, faculty reported that the resources provided were useful and they had good experiences working with peers. ✓,✗,-,n/a: same as Table 3. Multiple marks are present when the faculty member was teaching multiple courses.

| | P01 | P02 | P03 | P04 | P05 | P06 | P07 | P08 | P09 | P10 | P11 | P12 | P13 | P14 | P15 | P16 | P17 | P18 | Total |
|---|---|---|---|---|---|---|---|---|---|---|---|---|---|---|---|---|---|---|---|
| Backward design template useful? (Day 2) | ✓ | ✓ | - | ✓ | ✓ | ✓ | ✓ | n/a | ✓ | - | ✓ | ✓ | ✓ | ✓ | ✓ | - | - | n/a | 12/16 |
| Starter packs useful? (Day 2) | ✓ | ✓ | ✓ | ✓ | ✓ | ✓ | ✓ | n/a | - | ✓ | ✓ | ✓ | ✓ | ✓ | ✓ | - | ✓ | n/a | 14/16 |
| Online community used frequently? (Follow-Up #1) | ✓ | ✓ | - | - | ✓ | ✓ | ✓ | n/a | ✓ | ✓ | ✗ | - | ✗ | - | ✓ | - | - | n/a | 8/16 |
| Working with peers ok? (Day 3) | ✓ | ✓ | ✓ | ✓ | ✓ | ✓ | ✓ | n/a | ✓ | ✓ | ✓ | ✓ | ✓ | ✓ | ✓ | ✓ | ✓ | n/a | 16/16 |
| Integrated GenderMag concepts? (By first feedback) | ✓ | ✓ | ✓ | ✓ | ✓✓ | ✓ | ✓ | n/a | ✓ | ✓✓ | ✓ | ✓ | ✓ | ✓ | ✓✓✓ | ✓ | ✓ | ✓ | 21/21 |
| Materials fully approved? (By second feedback) | ✓ | ✓ | n/a | n/a | ✗✗ | ✓ | ✓ | n/a | ✓ | ✓✗ | ✗ | ✓ | ✓ | ✓ | ✓✗✗ | ✗ | ✓ | ✓ | 12/19 |
| Teaching GenderMag in course(s) this year | ✓ | ✓ | ✗ | ✗ | ✓ | ✓ | ✓ | ✗ | ✓ | ✓ | ✓ | ✓ | ✓ | ✓ | ✓ | ✓ | ✓ | ✓ | 15/18 |
| | | | | | | | | | | | | | | | | | | Overall success rate: 98/122 (80%) | |

## 4.4 Curriculum Element #4: Developing Faculty's Pedagogic Content Knowledge

Curriculum Element #4's Learning Outcome #4 aimed to help faculty prepare to engage and guide students on learning GenderMag concepts. As was mentioned in Section 2.4, education researchers have pointed out that knowledge of content is not enough to effectively *teach* that content; knowledge of how to effectively teach *this* content is called Pedagogic Content Knowledge (PCK). To help faculty harness and/or develop appropriate PCK for their additions, we used a combination of modeling, peer active learning, and prior PCK solutions for teaching inclusive design (Training of Trainers modeling of skills, skill practice, and feedback [6, 12, 33]; Community of Practice peer collaboration, support, and shared resources [48]):

---

Curriculum Element #4: PCK
- (1) See teaching of GenderMag concepts modeled
- (2) Practice teaching their new materials with peers
- (3) Known PCKs for teaching inclusive design

---

After seeing strategies for teaching GenderMag while learning it themselves during Day 1 (mechanism 1 above), faculty collaborated in small groups on developing each other's materials and took turns teaching the new content to each other (mechanism 2) for about four hours during the workshop. This work time was intended to help faculty practice teaching and find potential problems with their materials, promote collaboration, and provide an opportunity to complete much of their integration work during time they had already set aside. At first, their plans were so rough that problems were easily spotted. However, as they iterated, the peer playing the "student" role sometimes needed to deliberately act as though they were an uninterested, resistant, or obtuse student to bring out problems that can arise in those situations. (Some of the faculty were quite inventive in playing these roles.)

P06 (Follow-up 1): "Teaching <participant> a lesson we created together was incredibly fun and helpful…"

P12 (Follow-up 2): "It was productive to spend time hearing … how they taught the materials…"

P16 (Follow-up 2): "The breakout rooms were immensely helpful … in getting feedback on our initial ideas and how to improve upon them."

Once the faculty had helped one another unearth problems and attempt to resolve them, we introduced PCK elements from Oleson et al.'s field study of eight faculty members teaching inclusive design [30] (mechanism 3). From that study's 11 PCK practices, we emphasized four of them (Table 6). For example, our presentation of PCK-11 (Handling resistance) is shown in Table 6.

Curriculum Element #4's pairing of mechanisms (2) and (3) turned out to be quite effective. Although individual faculty's responses to questions asking about their confidence in teaching the material varied by timing and by content, 16/17 (94%) of reporting participants successfully reached the learning outcome, and the overall positive response rate was 86% (Table 7).

Table 6: (Left): The four PCKs highlighted from Oleson et al.'s study [30] on workshop Day 3.
(Right): Example: Excerpts from workshop slides on PCK11.

PCK1-Framing: Providing foundations first can give students the capacity to understand and engage with inclusive design methods

PCK2-Credibility: Providing students credible resources can convince them inclusive design methods are valid and useful

PCK6-TheoryOfMind: Coaching students to immerse themselves in the persona can help them with their "Theory of Mind" abilities to see software through the eyes of a persona

PCK11-HandlingResistance: Relating inclusive design methods' utility to the broader goal of inclusive appeal and/or greater market share can mitigate the risk of students' resistance and motivate them to learn inclusive design.

Table 7:

Curriculum #4 results from faculty responses to Day 1 and 3 Feedback as well as Follow up #2 about their ability to apply PCK and preparation to teach. Overall, participants indicated they were ready. ✓,✗,-,n/a: same as Table 3

| | P01 | P02 | P03 | P04 | P05 | P06 | P07 | P08 | P09 | P10 | P11 | P12 | P13 | P14 | P15 | P16 | P17 | P18 | Total |
|---|---|---|---|---|---|---|---|---|---|---|---|---|---|---|---|---|---|---|---|
| Can answer student questions about GenderMag eval? (Feedback Day 1) | ✓ | ✓ | ✓ | - | ✓ | ✓ | ✓ | ✗ | ✓ | ✓ | ✓ | ✓ | ✓ | ✓ | ✓ | ✓ | ✓ | n/a | 15/17 |
| Prepared to apply PCK? (Feedback Day 3) | ✓ | ✗ | ✓ | ✓ | ✓ | ✓ | - | n/a | - | ✓ | ✓ | ✓ | ✓ | ✓ | ✓ | ✓ | ✓ | n/a | 13/16 |
| Teach GenderMag in engaging way? (Follow-up #2) | ✓ | ✓ | - | ✓ | ✓ | ✓ | ✓ | n/a | ✓ | ✓ | ✓ | ✓ | ✓ | - | ✓ | ✓ | ✓ | n/a | 14/16 |
| | | | | | | | | | | | | | | | | Overall success rate: | | 42/49 (86%) | |

## 4.5 Faculty Reflections

Up to this point, we have evaluated the educate-the-educators curriculum's effectiveness by each of their learning outcomes. As Tables 3, 4, 5, and 7 show, learning outcome rates for each curriculum element ranged from 71% to 94%. Now, we consider how these learning outcomes translated to the experiences of the ten faculty whose revised courses occurred in the fall term schedule. These 10 faculty taught 16 sections of 7 courses covering CS0, CS1, CS2, Web Design, Mobile Apps, Software Engineering, and Education (non-CS). In total, they taught 275 students.

After teaching the inclusive design content, each faculty member shared their teaching experiences through the interviews we conducted at the end of fall term (Section 3) and we compared these interview results with the learning outcome results from Sections 4.1-4.4.

Most of the faculty's interview content indicated that their fall courses went at least as well as the faculty had expected (Table 8). In fact, 91% of these findings agreed with or were more positive than the expectations they had reported earlier (Tables 3, 4, 5, and 7). Table 8 summarizes their positive and negative reflections upon how the term went for them as compared with the curriculum element evaluations of the previous four sections.

Table 8: Fall Term faculty reflections.  +: positive reflection, -: negative reflection, ~: indirect positive reflection.
Comparisons with Tables 3, 4, 5, and 7: green: better than anticipated, yellow: same as anticipated, orange: worse than anticipated.
(Note: P18 did not fill out the questionnaires for Table 3, Table 4, Table 5, Table 7.)

| Curriculum Element | | P01 | P05 | P07 | P09 | P10 | P11 | P15 | P16 | P17 | P18 |
|---|---|---|---|---|---|---|---|---|---|---|---|
| 1: Motivating | Burden to integrate light? | + - | + | + | + | + | + - | | | + | - |
| | Material relevant to your students? | | | + | + | | | + | | + | + |
| | Benefit of incorporating into course? | + | | + | + | | + | | + | | + |
| 2: Content | Can do GenderMag  eval? | | | + | | | | + | | | +~ |
| | GenderMag found meaningful issues? | | | | | | | | | | + |
| 3: Embedding | Online community used? | | | + | | | | | | | |
| 4: PCK | Can answer student questions about GenderMag? | + - | - | | + | | | + | | | |
| | Prepared to apply PCK? | + | | + | + | | + | + | + | + | + |
| | Teach GenderMag in engaging way? | + | | | ~ | | ~ | | + ~ | ~ | ~ |

Relating to Curriculum Element #1 (Motivating), faculty reflected upon the costs and benefits they had experienced when they taught the content to their students. Most faculty were positive about the burden to integrate the content, and three faculty members said it was lighter than they had expected. However, a few faculty members experienced a burden they regarded as heavy. For P18, the burden was due to grading and collecting data:

> P18 (Interview): "I think it went well. Unfortunately, I'm transitioning back to face-to-face myself, and I think I created a very poor assignment, which, as a result, I had to spend quite a bit of time grading…"

Nonetheless, they seemed optimistic about addressing the roadblocks in future terms:

> P18 (Interview): "…the <course> is taught in spring so I'll be doing it there… and then I would certainly do it again next fall in <course> and again make sure I'm getting the information I want without <making grading and data collection hard>."

The remaining two faculty reporting burdens described front-end (course planning) burdens, namely, finding time in the course for added material, especially given the context of the pandemic and the return to in-person learning:

> P01 (Interview): "I don't have enough time <in the course>…because of the transition from online to non-online. Some <students were> quarantined, some students <tested> positive..."

> P11 (Interview): "…there's a lot of material to cover and…it's very hard to squeeze…in all of <the regular content> and also this."

Still the same two faculty members did not see actually running their revised courses as burdensome, once they had figured out their changes:

> P01 (Interview): "So it's really not much difference <in workload>… it's not a problem for me at all."

> P11 (Interview): "No, <the only workload is> just the extra assignment to grade… It's easy. It's not difficult"

The bottom line for P11 was that the benefits outweighed the costs. In fact, they advocated for growing the effort.

> P11 (Interview): "Cost is, I guess, maybe you have to go through the workshop and understand and stuff but... I think the cost is minimal and the effect much overweighs it."

> P11 (Interview): "I think we could get all the other faculty kind of on board so that you know they could try it out."

In total, 7 of the 10 interviewees spoke of the benefits and/or relevance of adding this content to their courses.

> P17 (Interview): "I think that the students realizing <which persona> they directly related to, I think is a great thing to walk away from here."

P16 (Interview): "I think opening their minds to other perspectives…there's obviously a great need for more of it and I think the students did get that through the experience <with inclusive design content>."

Interviewees offered few remarks that related to the learning outcomes of Elements #2 (Content) and #3 (Embedding), but the points these faculty made about these topics aligned with the earlier learning outcome evaluations. One example was P18's description of an instance in which the new content helped students who were developing a clothing retail application to find meaningful issues (Element #2's learning outcome):

P18 (Interview): "And at first they had thought that people would go there with the demographic of say 18 to 22 and buy clothing. But when they started working with the GenderMag, they suddenly realized people might go to buy clothing for people 18 to 20 and they had not thought about, you know, the super-buyer."

The interviewees had a great deal to say about Curriculum Element #4 (PCK). Most faculty felt reasonably successful in this regard. Compared to the workshop, two faculty members were able to demonstrate use of the PCK after previously indicating that they were neither prepared nor unprepared to do so. However, two faculty members also reported being unable to answer some of the questions that arose in their classes.

P01 (Interview): "<Some students> think that <Computer Science is> equal already; they shouldn't need to talk about <GenderMag>... I did not <respond>."

P05 (Interview): "<The students> felt like if the site was accessible… we didn't need to apply the GenderMag method, because we're thinking of all these accessibility use cases… I didn't have a follow up to that."

On the other hand, two others reported being comfortable answering the students' questions.

P16 (Interview): "At least two students that were like, 'Why are we doing this? Like what are we supposed to be doing?' And so of course, I responded and really emphasized why this is important for web developers as they enter the profession to really be familiar with."

P09 (Interview): "Once I went to your website and I explained what it really is [using] some charts and diagrams on the website, too, nobody had any follow-up questions."

One faculty member reported that some students were overwhelmed by too much overlap in the content among different courses, which was a situation we had been trying to avoid through the collaboration/community aspects of the curriculum:

P05 (Interview): "I know a couple…students <who are> doing <inclusive design content> for all of their other classes and they are just kind of burnt out over it. I think it's…because…new stuff coming at them from several different professors, several different ways of doing it."

In contrast, another faculty member reported some students were interested in learning more:

P01(Interview): "<There was> somebody enthusiastic and they say they want to know more. I had to tell them that you have to take the HCI class which is elective."

These results will inform future iterations on our educate-the-educators curriculum and the accompanying resources we provide to support it. For example, we could specifically train faculty to answer common questions and use the PCK to address concerns. In fact, P05 directly asked for a review of the workshop material and references for answering common questions. As we continue to adapt our online community, we hope to make these resources readily available and easily accessible.

Overall, faculty reported that the majority of student reactions were positive and engaged.

P17 (Interview): "The good reaction I thought was when my students said that they directly relate it to the persona."

P16 (Interview): "…it was interactive, it was engaging. The students were lively"

P01 (Interview): "Unexpectedly, people talked so much about <the inclusive design personas> and then made the final project more lively…they <were> more active in doing <the> project"

## 5 CONCLUDING REMARKS

In this paper, we have presented an educate-the-educators curriculum to enable faculty to carry out a coordinated, four-year vision of graduating computing students who do inclusive design as part of their everyday software development work. We built this educate-the-educators curriculum "on the shoulders of giants" by following recommendations from earlier work from the accessibility education community, Community of Practice research, Training of Trainers research, and Pedagogic Content Knowledge results from teaching inclusive design.

We investigated the approach's efficacy over a three-month period and end-of-fall interviews, in which 18 faculty experienced our educate-the-educators curriculum and decided whether and how to follow up. Ten of the faculty then took their GenderMag-integrated course materials into their classrooms fall term. (The other faculty's courses are scheduled in future terms.) We followed their progress from the outset of the project through their end-of-summer preparations to offer the materials they had developed to their classes, and then gathered how their changes had turned out in their classrooms.

Among the results were:

- Curriculum Element #1 (Motivating the faculty): Included in the questions we considered were whether faculty would choose to participate in a coordinated effort to embed inclusive design across the four-year curriculum, whether they would find the time cost of doing so to be onerous, and whether they foresaw benefits to their courses and their students from doing so. The field study results showed that none of the faculty considered the cost to be heavy, and almost all of them saw the approach to be beneficial to their courses. Although one participant dropped out at this stage, 14/17 (82%) of reporting faculty met the learning outcome threshold. Overall, 35/49 (71%) of faculty responses were favorable, only 2/49 (4%) were negative, and the remaining responses were neutral (from Table 3).

- Curriculum Element #2 (Teaching the faculty inclusive design via GenderMag): This curriculum element was a prerequisite for the rest of the project. The challenge was to enable faculty across a wide range of CS specialties to understand inclusive design well enough to teach portions of it to their classes. Fortunately, the results were very positive: 16/17 of the faculty reported the method to be useful, and 14/17 reported being able to perform GenderMag evaluations, and 14/17 (82%) of reporting faculty met the learning outcome threshold. In total, 30/34 (88%) of the faculty responses were favorable (Table 4).

- Curriculum Element #3 (Embedding GenderMag concepts into their courses): At this point, faculty needed to act upon what they had learned, by starting work on integrating inclusive design concepts into their courses. This curriculum element made extensive use of collaboration among small peer groups, personalized coaching, and several feedback loops. Along the way, two faculty eventually decided that their courses could not profitably include the material, but the remainder followed through, ultimately preparing all the materials they would need to walk into class this fall. 13/18 (72%) of reporting faculty met the learning outcome threshold. Overall, responses for this curriculum element were 98/122 (80%) positive (Table 5).

- Curriculum Element #4 (Developing faculty's PCK): This curriculum element aimed to help faculty progress from just teaching the new materials to teaching them *effectively*. Ultimately, faculty rated their abilities in this area positively: almost all reported being prepared to answer students' questions about the material, teach the material engagingly, and apply the PCK elements they had found useful. 16/17 (94%) of the reporting faculty met the learning outcome threshold, and overall this curriculum element received 42/49 (86%) positive responses (Table 7).

- Results from the end-of-fall interviews: Ten faculty members' updated classes occurred during fall term, in which they taught 16 sections of 7 courses, which served a total of 275 students. 91% of their reported outcomes indicated that their incorporations of inclusive design concepts into their courses went as well as or better than they had expected (Table 8).

As to the "big picture", how well did the faculty's efforts fit together? Table 9 overviews the year-by-year coverage of the inclusive design content that faculty implemented in their courses. As the table shows, the most basic level of inclusive design (understanding differences in cognitive styles) is introduced in Year 1 and applied across all four years. Over the next years, the level of student engagement in inclusive design increases, until by Year 4 they are applying the full GenderMag method to their coursework.

Table 9: By Fall 2021, GenderMag inclusive design concepts were integrated throughout the 4-year undergraduate CS/IT curriculum at University X. (Light): Learning; (Dark): After learning, acting upon. Year 1: CS0; Year 2: CS1; Year 3: Web Dev., Project Management, Object Oriented Design, Mobile Apps., HCI; Year 4: Software Eng., Capstone.

| | Year 1 | Year 2 | Year 3 | Year 4 |
|---|---|---|---|---|
| (Basic inclusion) Cognitive styles | | | | |
| (Apply to software) Evaluation | | | | |
| (Act upon evaluation) Fixes | | | | |
| (Systematic process) Full GenderMag method | | | | |

These results are very encouraging, but we have investigated them only in the context of a primarily undergraduate institution, where the primary goal is quality teaching. Other four-year college settings in which quality teaching is a critical component of the faculty reward system may be able to carry out this curriculum with very little adjustment. Two-year college settings such as community colleges might be able to use all four elements, but with a less populated "big picture" (recall Figure 3). One open question is how Ph.D.-granting universities can tailor Curriculum Element #1 to their university's reward system for research-active faculty; one possibility may be to tie it to their existing BPC efforts to broaden participation in computing. More generally, the question of how our results generalize to other kinds of settings remains open. Still, we believe our four educate-the-educators curriculum elements, and the mechanisms to carry them out, are generalizable to other contexts. We hope other researchers will join us in helping to find out.

## ACKNOWLEDGMENTS

Anonymized for blind review.